\documentclass{ptephy_v1-1912.12229}

\preprintnumber{arXiv:1912.12229} 

\usepackage{amsmath,amssymb,revsymb,graphicx,dcolumn} 

\begin{document}

\title{On the emergence of an expanding universe
from a Lorentzian matrix model}

\author{F.R. Klinkhamer}
\affil{Institute for Theoretical Physics,
Karlsruhe Institute of Technology (KIT),\\
76128 Karlsruhe, Germany \email{frans.klinkhamer@kit.edu}}

\begin{abstract}
We present evidence that recent numerical results from
the reduced classical equations of the Lorentzian IIB matrix model
can be interpreted as corresponding to the emergence of an
expanding universe. In addition, we propose an effective
metric to describe the emerging (3+1)-dimensional spacetime.
This metric gives, at all times,
finite values for the Ricci and Kretschmann curvature scalars.
With these results, we are able to
give a heuristic discussion of the origin of the Universe
in the context of the IIB matrix model.
\end{abstract}

\subjectindex{B25,\, B83 \vspace*{-7mm}}  

\maketitle

\section{Introduction}
\label{sec:Intro}

The Ishibashi--Kawai--Kitazawa--Tsuchiya (IKKT)
large-$N$ matrix model~\cite{IKKT-1997} has been proposed
as a nonperturbative
definition of (9+1)-dimensional type--IIB superstring theory.
The model is also known as the IIB matrix model and
further discussion appears in Ref.~\cite{Aoki-etal-review-1999}.

Numerical
investigations~\cite{KimNishimuraTsuchiya2012,NishimuraTsuchiya2019}
of the Lorentzian version of the IKKT matrix model
have found indications that three spatial dimensions emerge
which behave differently from the remaining six spatial dimensions
and that the ``late-time'' behavior of these three spatial dimensions
may be controlled by a classical configuration of the matrices.
Recently, further numerical results have been
presented~\cite{Hatakeyama-etal2020}
from the matrix-model classical equations with
a particular \textit{Ansatz} for the ten bosonic matrices,
where the \textit{Ansatz}
implements the split of a (3+1)-dimensional ``spacetime''
and a  6-dimensional ``internal space.''

The goal here is to present a simple check that Figs. 3--5
in Ref.~\cite{Hatakeyama-etal2020}
indeed correspond to a flat 3-space expanding with cosmic time $|t|$.
In addition, we suggest an effective
metric for the emerging (3+1)-dimensional spacetime
and give a heuristic discussion of the origin of the Universe,
assuming that the matrix model describes a new phase
replacing the big bang
singularity~\cite{Klinkhamer2019,Klinkhamer2020,KlinkhamerWang2019}.
We have also obtained certain exact solutions
of the reduced classical equations of the Lorentzian matrix model,
which are presented and discussed in 
Appendix~\ref{app:Exact-solutions}.   
Incidentally, the adjective ``classical'' is to be used
with caution, as will be explained in Sect.~\ref{sec:Discussion}.

\section{Procedure}
\label{sec:Procedure}

The basic idea is to calculate certain observables
from a simple model describing
random points inside a box embedded in Euclidean 3-space
and to compare them with the observables obtained from the
reduced classical equations of the Lorentzian matrix
model~\cite{Hatakeyama-etal2020}.
The procedure consists of six steps:
\begin{enumerate}
\item
Consider a cube of volume $(2\,L)^{3}$
in $\mathbb{R}^{3}$ with a constant flat metric,
$g_{ij}(x)=\delta_{ij}$ for spatial indices $i,\,j \in \{1,\, 2,\, 3\}$.
\vspace*{-.0mm}
\item
Generate $n$ points in the cube with random (uniform-distribution) values
for the Cartesian coordinates $x^{i}$ in the range $[-L,L] \subset \mathbb{R}$.
\vspace*{-.0mm}
\item
Make diagonal $n \times n$ matrices $\widehat{X}^{1}$, $\widehat{X}^{2}$,
and $\widehat{X}^{3}$, where
the matrix $\widehat{X}^{1}$ has the $x^{1}$ coordinates
of the $n$ random points on the diagonal and zeros elsewhere,
and similarly for the matrices $\widehat{X}^{2}$ and $\widehat{X}^{3}$.%
\vspace*{-.0mm}
\item
Define the $3 \times 3$ matrix $\widehat{T}$ as in Eq.~(3.5)
of Ref.~\cite{Hatakeyama-etal2020} and calculate its
eigenvalues $\widehat{\lambda}_i\,$.
\vspace*{-.0mm}
\item
Define the $n \times n$ matrix $\widehat{Q}$ as in Eq.~(3.7)
of Ref.~\cite{Hatakeyama-etal2020} and calculate its eigenvalues $\widehat{q}_r$,
for $r=1,\, \ldots \,,\, n$.
\vspace*{-.0mm}
\item
Define the number $\widehat{R}^{\,2}$
as in Eq.~(3.8) of Ref.~\cite{Hatakeyama-etal2020} and calculate its value.
\end{enumerate}
The explicit definitions of 
$\widehat{T}$, $\widehat{Q}$, and $\widehat{R}^{\,2}$ are:  
\begin{subequations}\label{eq:That-def-Qhat-def-Rhatsquare-def}
\begin{eqnarray}
\label{eq:That-def}
\big(\;\widehat{T}\;\big)^{ij} &\equiv&
\frac{1}{n}\, \text{tr}\left[\widehat{X}^{i} \cdot \widehat{X}^{j}\right]\,,
\\[2mm]
\label{eq:Qhat-def}
\widehat{Q} &\equiv& \delta_{ij}\,\widehat{X}^{i} \cdot \widehat{X}^{j}\,,
\\[2mm]
\label{eq:Rhatsquare-def}
\widehat{R}^{\,2}
&\equiv&
\frac{1}{n}\,\sum_{r=1}^{n}\,\widehat{q}_r\,,
\end{eqnarray}
\end{subequations}
where the central dot on the right-hand sides of
\eqref{eq:That-def} and \eqref{eq:Qhat-def}
stands for matrix multiplication.
The flat metric $\delta_{ij}$ in \eqref{eq:Qhat-def}
effectively traces back to the space--space components of the
coupling constants $\eta_{KL}$ in the Lorentzian version of the
IKKT matrix model~\cite{IKKT-1997,KimNishimuraTsuchiya2012},
using the notation of Ref.~\cite{Hatakeyama-etal2020}
with (9+1)-dimensional indices $K$ and $L$
running over $\{0,\, 1,\, 2,\, 3,\, \ldots \,,\, 9\}$.

The observables obtained from the random-point model
carry a hat, in order to distinguish them from the observables
obtained from the classical equations of the matrix
model~\cite{Hatakeyama-etal2020}.

\section{Numerical results}
\label{sec:Numerical-results}

The dimensionless length scale $L$ of the box used in the procedure
of Sect.~\ref{sec:Procedure}
can be considered to be a function of the absolute value of the
dimensionless  cosmic time $t$,
\begin{equation}\label{eq:L-of-ABSt}
L=L\big(|t|\big)\,.
\end{equation}
The $n$ random points from Sect.~\ref{sec:Procedure}
are then generated inside a cube of volume
\begin{equation}
\label{eq:V-cube}
V_\text{cube}(t)=
\big[2\,L(|t|)\big] \times \big[2\,L(|t|)\big] \times \big[2\,L(|t|)\big]\,.
\end{equation}
For the moment, $t$ is considered to be just an arbitrary
cosmic-time coordinate, but $t$ has, in fact, been found
to arise from the matrix-model calculation of
Ref.~\cite{Hatakeyama-etal2020}
(see Sect.~\ref{sec:Matrix-model-configurations} below for further discussion).

\begin{figure}[t]
\vspace*{-0mm} 
\begin{center} 
\includegraphics[width=1.00\textwidth]{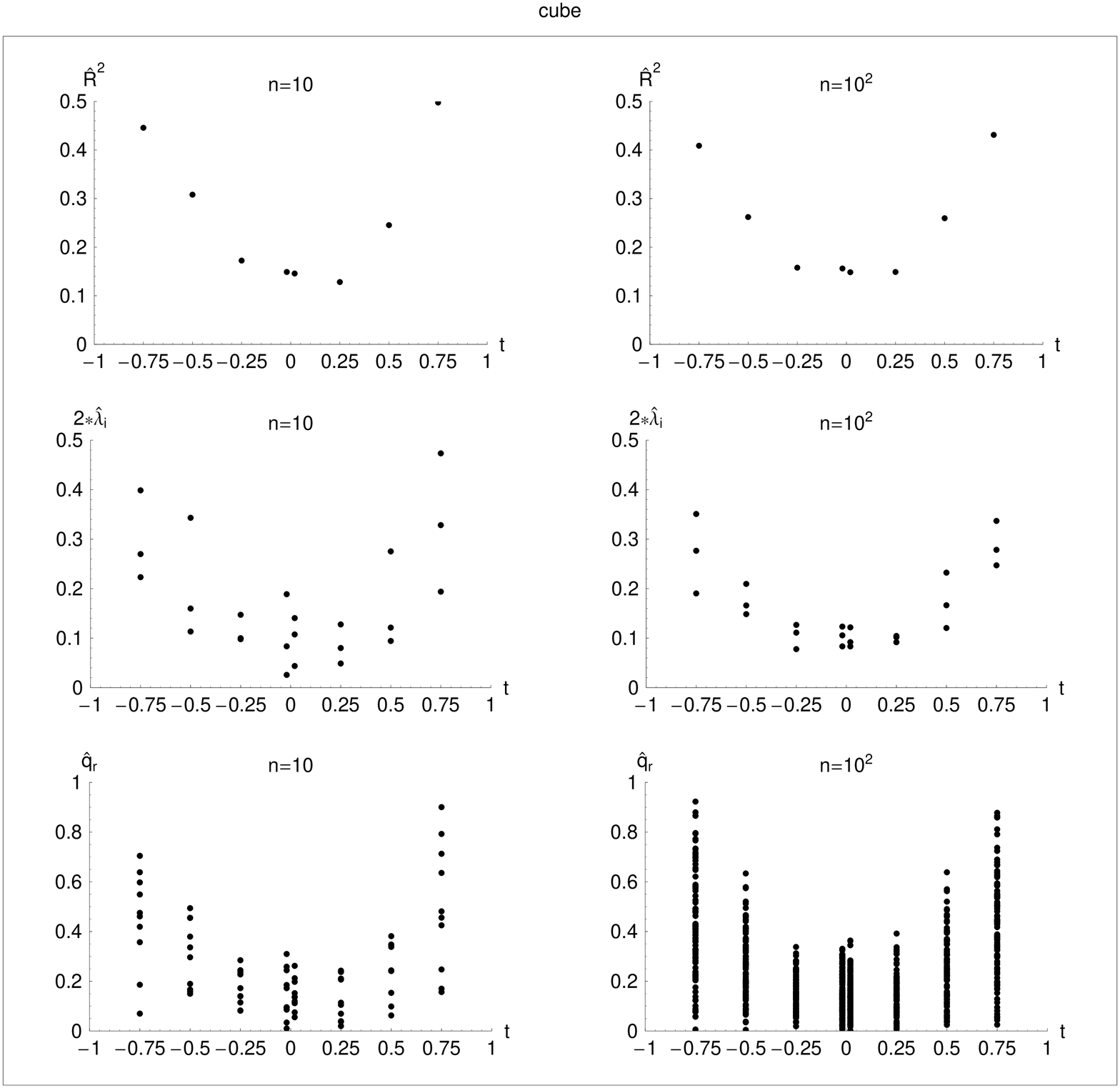}
\end{center}
\vspace*{-0mm}
\caption{Top row: Numerical results for the quantity
$\widehat{R}^{\,2}$ as defined by \eqref{eq:Rhatsquare-def} from a typical run
with a cube of volume \eqref{eq:V-cube} for
length scales \eqref{eq:L-of-t-values}.
The procedure used for generating the diagonal $n \times n$
matrices $\widehat{X}^{i}$ involves
$n=10$ (left) or $n=10^{2}$ (right) random points in a cube
and is outlined in Sect.~\ref{sec:Procedure}.
Middle row: Corresponding numerical results
for the eigenvalues $\widehat{\lambda}_i$ of the matrix $\widehat{T}$
as defined by \eqref{eq:That-def}, where actually twice the eigenvalues
are plotted.
Bottom row: Corresponding numerical results
for the eigenvalues $\widehat{q}_r$ of the matrix $\widehat{Q}$
as defined by \eqref{eq:Qhat-def}.}
\label{fig:fig1}
\vspace*{0mm}
\end{figure}

We start by taking $n=10$ random points
(in order to allow for a direct comparison with the results
of Ref.~\cite{Hatakeyama-etal2020}; see below)
and assume the following numerical values for the box scale $L(|t|)$:\vspace*{-0mm}
\begin{equation}\label{eq:L-of-t-values}
L(0.02) = 0.4\,,
\quad
L(0.25) = 0.4\,,
\quad
L(0.50) = 0.5\,,
\quad
L(0.75) = 0.65\,,
\end{equation}
where the actual $L$ values are chosen to obtain
the desired values for $\widehat{R}^{\,2}$.
In fact, the values for $\widehat{R}^{\,2}$  obtained in a typical run
are shown in the top-left panel of Fig.~\ref{fig:fig1}.
The corresponding eigenvalues $\widehat{\lambda}_i$
are shown in the middle-left panel of Fig.~\ref{fig:fig1}
and the corresponding eigenvalues $\widehat{q}_r$ in
the bottom-left panel of Fig.~\ref{fig:fig1}.
Two remarks are in order. First, with the small number
of points ($n=10$), the scatter of $\widehat{\lambda}_i$
values at $|t|=0.75$ is significant.
Second, in the bottom-left panel of Fig.~\ref{fig:fig1},
the apparently increasing gap near zero for larger and larger values of
$|t|$ just appears because a larger and larger interval
is covered by a fixed number ($n$) of points.
Both issues will be discussed further below.

\begin{figure}[t]
\vspace*{-0mm}
\begin{center}
\includegraphics[width=1.00\textwidth]{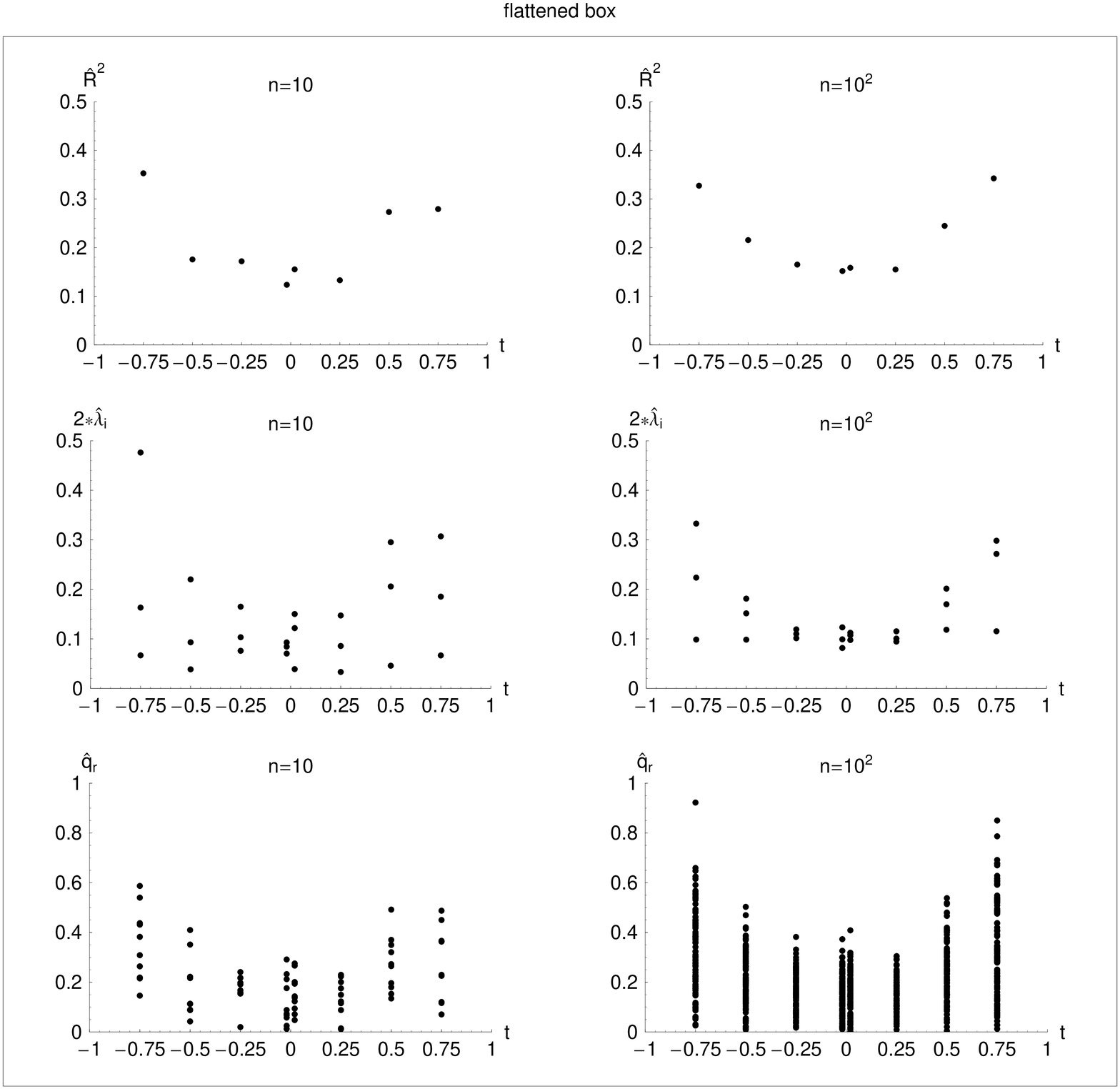}
\end{center}
\vspace*{-0mm}
\caption{Same as Fig.~\ref{fig:fig1}, but now for $n$ random points in a
flattened box of volume \eqref{eq:V-flattened-box} for
length scales \eqref{eq:L-of-t-values}.}
\label{fig:fig2}
\vspace*{0mm}
\end{figure}

For a first comparison,
we put $n=10$ random points in a flattened box of volume
\begin{equation}
\label{eq:V-flattened-box}
V_\text{flattened-box}(t)=
\big[2\,L(0)\big] \times \big[2\,L(|t|)\big] \times \big[2\,L(|t|)\big]\,,
\end{equation}
with $L(0)=L(0.02)$ from \eqref{eq:L-of-t-values}.
The corresponding numerical results are shown in
the left panels of Fig.~\ref{fig:fig2}.
The middle-left panel of Fig.~\ref{fig:fig2}
shows that two eigenvalues ($\widehat{\lambda}_{1,2}$)
grow as $|t|$ increases, whereas one eigenvalue ($\widehat{\lambda}_{3}$)
remains more or less constant.
The behavior in the middle-left panel of Fig.~\ref{fig:fig2}
is fundamentally different from that
in the middle-left panel of Fig.~\ref{fig:fig1}.

\begin{figure}[t]
\vspace*{-0mm}  
\begin{center}
\includegraphics[width=1.00\textwidth]{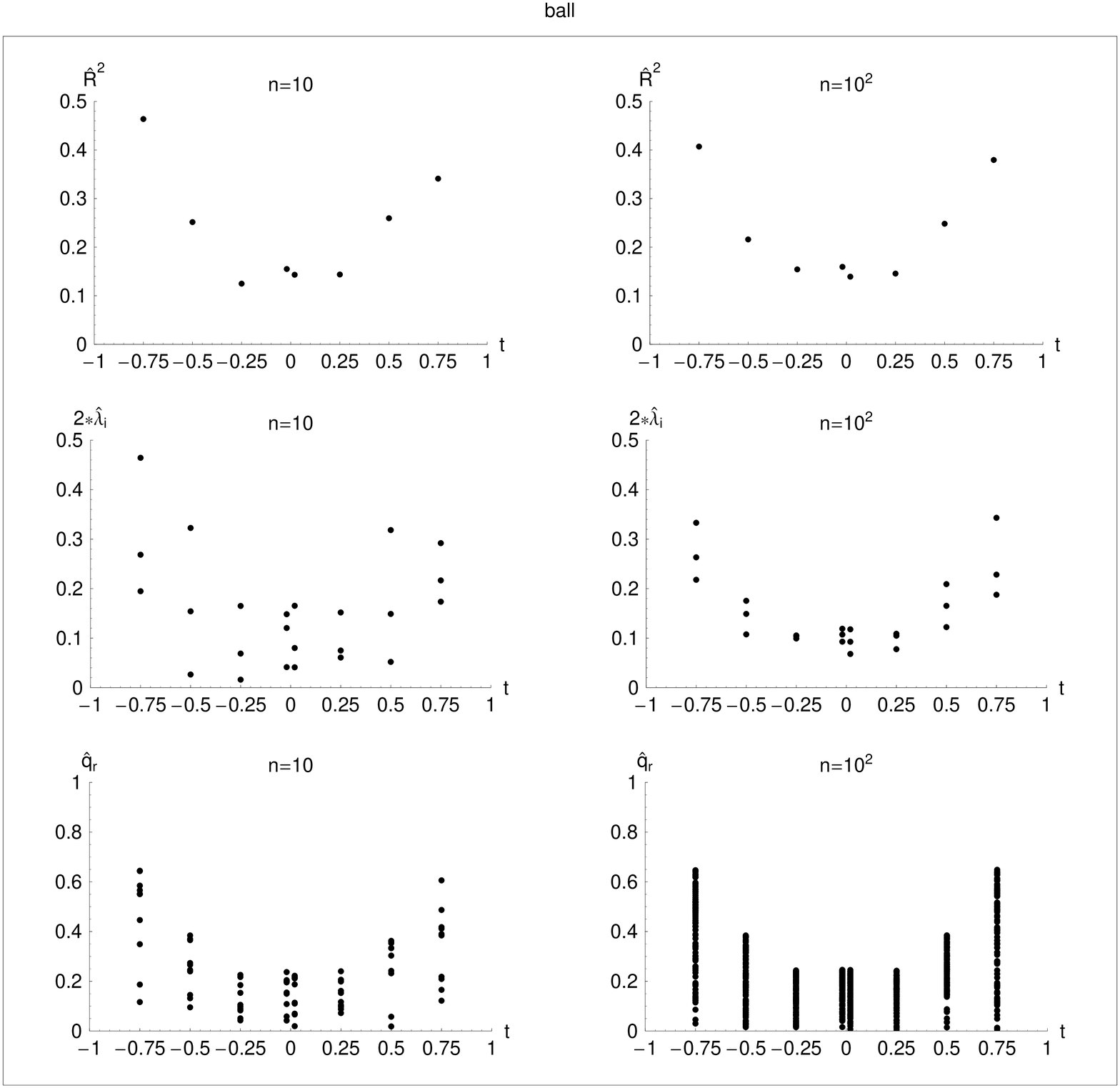}
\end{center}
\vspace*{-0mm}
\caption{Same as Fig.~\ref{fig:fig1}, but now for $n$ random points in a
ball of volume \eqref{eq:V-ball} for length scales \eqref{eq:L-of-t-values}.}
\label{fig:fig3}
\vspace*{0mm}
\end{figure}

For a second comparison,
we put $n=10$ random points in a ball of
radius $\widetilde{L}(t)=(6/\pi)^{1/3}\,L(t)$
and volume
\begin{equation}
\label{eq:V-ball}
V_\text{ball}(t)=
(4\pi/3)\,\Big[\widetilde{L}(t)\Big]^{3} = 8\,\big[L(t)\big]^{3}\,.
\end{equation}
The corresponding numerical results are shown in
the left panels of Fig.~\ref{fig:fig3}, which are qualitatively
the same as those in  the left panels of Fig.~\ref{fig:fig1}
(but there are differences, as will become clear later on).

\begin{figure}[t]
\vspace*{-0mm}
\begin{center}
\includegraphics[width=1.00\textwidth]{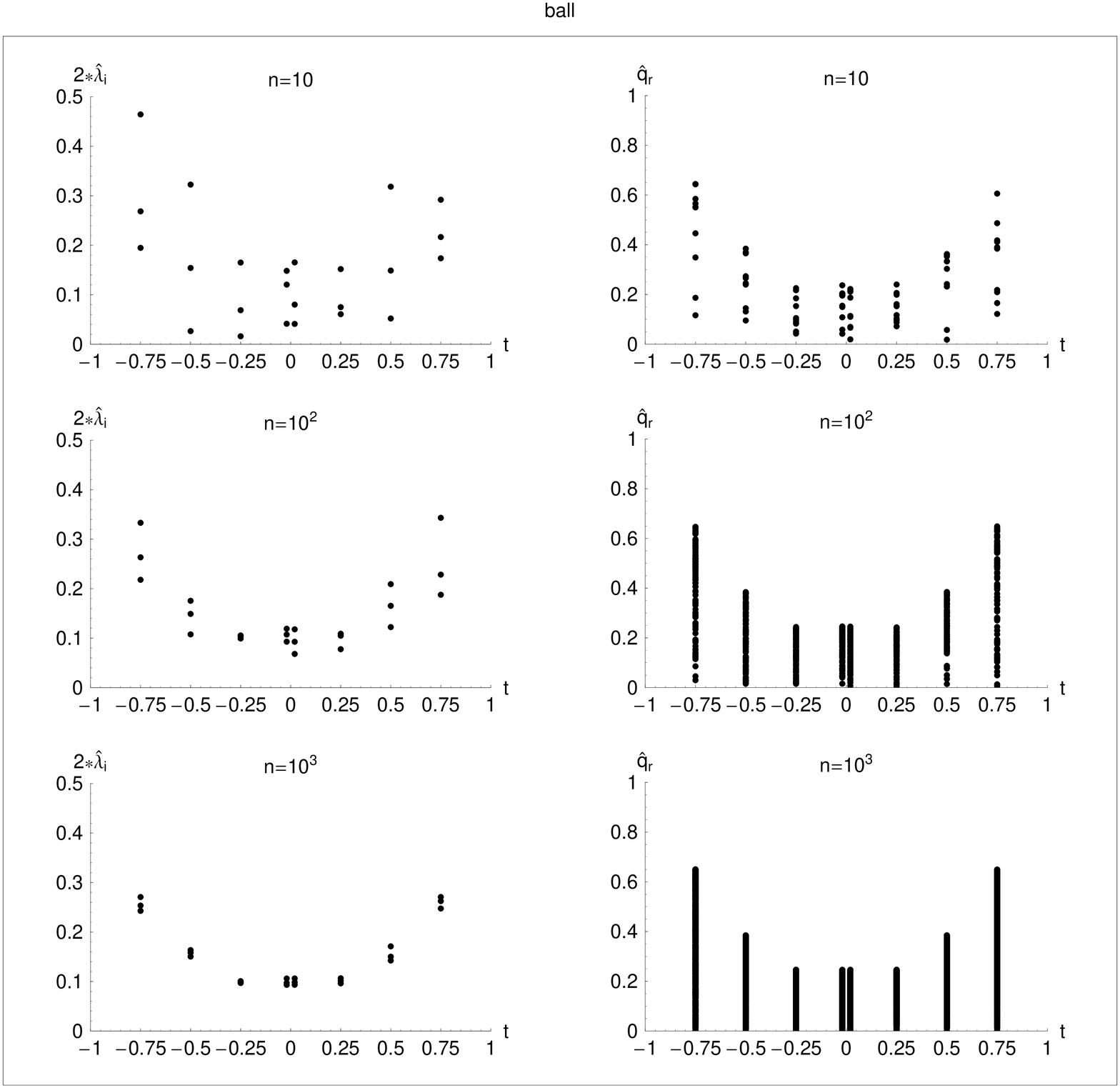}
\end{center}
\vspace*{-0mm}
\caption{Numerical results from Fig.~\ref{fig:fig3}, together with
numerical results from $n=10^{3}$ random points.
The left panels show the reduction of the
spread of $\widehat{\lambda}_i$ eigenvalues as $n$ is increased,
while the right panels show the augmentation of the density of
$\widehat{q}_r$ eigenvalues as $n$ is increased.
}
\label{fig:fig4}
\vspace*{0mm}
\end{figure}

\begin{figure}[t]
\begin{center} 
\includegraphics[width=1.00\textwidth]{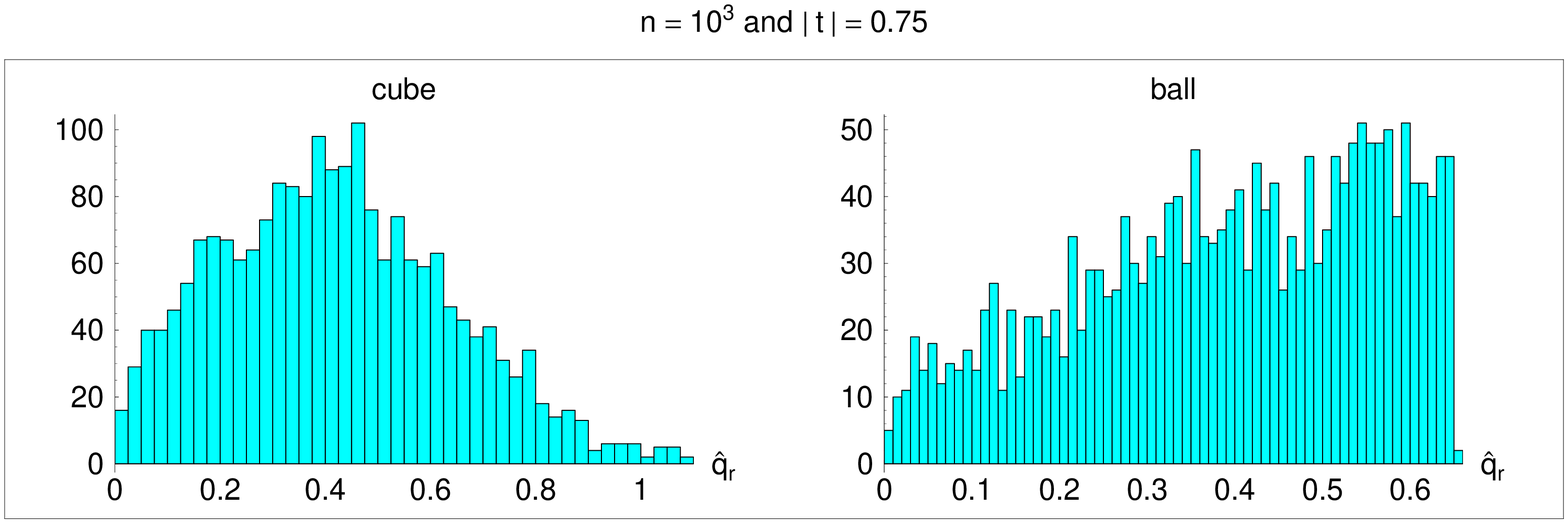}
\end{center}
\caption{Distribution of $\widehat{q}_r$ eigenvalues for
the case of the cube (left) and the case of the ball (right), with
$n=10^{3}$ random points and at cosmic time $|t|=0.75$.}
\label{fig:fig5}
\vspace*{0mm}
\end{figure}

With the simple random-point model of Sect.~\ref{sec:Procedure}, it is
easy to push the numerics further, for example,  
by taking $n=10^{2}$ random points
(the results from $n=10^{3}$ random points are similar and
a selection will be given later).
The numerical results from $n=10^{2}$ random points
are given by the right panels of
Figs.~\ref{fig:fig1}--\ref{fig:fig3}.
The top-right and middle-right panels of
Figs.~\ref{fig:fig1} and \ref{fig:fig3} show a more or
less stable behavior of $\widehat{R}^{\,2}(t)$  and $\widehat{\lambda}_i(t)$
\big[for $n\to\infty$ at fixed $t$, the eigenvalues $\widehat{\lambda}_i$
approach each other
and $\widehat{\lambda}_i \sim \widehat{R}^{\,2}/3\,$\big],
while the bottom-right panels of
Figs.~\ref{fig:fig1} and \ref{fig:fig3} show a dense distribution
of $\widehat{q}_r$ over a range increasing with $|t|$.
For $n=10^{2}$ random points in the flattened box,
the middle-right panel of Fig.~\ref{fig:fig2} gives results similar
to those for $n=10$ in the middle-left panel,
with two eigenvalues $\widehat{\lambda}_{1,2}(t)$   
growing with increasing $|t|$ and a single eigenvalue 
$\widehat{\lambda}_{3}(t)$  staying more or less constant
(the same behavior is seen with $n=10^{3}$ random points in the flattened box).

We have already mentioned the significant spread of $\widehat{\lambda}_i$
values if the number of points is relatively small ($n=10$)
and the gaps occurring in the $\widehat{q}_r$ distribution.
Figure~\ref{fig:fig4} shows, for the case of a ball,
the behavior with increasing values of $n$.
Recall that the quantity $\widehat{R}^{\,2}$ is given by the
sum of the three eigenvalues $\widehat{\lambda}_i$ or,
alternatively, by the average of the eigenvalues $\widehat{q}_r$,
according to \eqref{eq:Rhatsquare-def}.
As to the detailed distribution of $\widehat{q}_r$
eigenvalues, Figs.~\ref{fig:fig1} and \ref{fig:fig3}
show some differences between the case of the cube and
the case of the ball. Figure~\ref{fig:fig5} gives the
corresponding distributions of $\widehat{q}_r$ values
for $n=10^{3}$ random points and cosmic time $|t|=0.75$.

To summarize, we have found
that the left panels of Figs.~\ref{fig:fig1} and \ref{fig:fig3}   
from the random-point-model calculations
directly match Figs.~3--5 from the matrix-model
calculation~\cite{Hatakeyama-etal2020}.
The conclusion is that the classical matrix-model
equations (2.6) of Ref.~\cite{Hatakeyama-etal2020}
with the appropriate \textit{Ansatz} (2.7)
of Ref.~\cite{Hatakeyama-etal2020}
indeed seem to generate a cosmic time $t$
and a flat 3-space that expands with $|t|$.

\section{Effective metric}
\label{sec:Effective-metric}

Taking the classical results from Ref.~\cite{Hatakeyama-etal2020}
at face value and relying on the spacetime interpretation from
Sect.~\ref{sec:Numerical-results}, we postulate the following effective
metric~\cite{Klinkhamer2019,Klinkhamer2020}:%
\begin{subequations}\label{eq:mod-FLRW}
\begin{eqnarray}\label{eq:mod-FLRW-ds2}
\hspace*{-2mm}
ds^{2}\,\Big|^\text{\,(eff)}
&\equiv&
g_{\mu\nu}^\text{\,(eff)}(x)\, dx^\mu\,dx^\nu
=
- \frac{t^{2}}{t^{2}+b^{2}}\,dt^{2}
+ a^{2}(t) \;\delta_{ij}\,dx^{i}\,dx^{j}\,,
\\[2mm]
\hspace*{-2mm}
b &>& 0\,,
\\[2mm]
\hspace*{-2mm}
t &\in& (-\infty,\,\infty)\,,\quad
x^{i} \in (-\infty,\,\infty)\,,
\end{eqnarray}
\end{subequations}
where relativistic units are used ($c=1$) and
where $b$ can be interpreted as the length scale
of a spacetime defect~\cite{Klinkhamer2019-JPCS}.
The individual components
of $x^\mu =\big( c\,t,\,x^{1},\,x^{2},\,x^{3} \big)$
have the dimension of length.

The square of the cosmic scale factor $a(t)$
in the metric \eqref{eq:mod-FLRW} can be identified with the
quantity $R^{\,2}(t)$ obtained from the classical results of the
large-$N$ matrix model (see Fig.~3 of Ref.~\cite{Hatakeyama-etal2020}).
Concretely, we take
\begin{equation}
\label{eq:a-of-t-from-matrix-model-R2}
a(t) =
L(t) \approx 0.4\,\big(1+t^{2}/b^{2}\,\big)\,,
\end{equation}
which reproduces approximately the values of
\eqref{eq:L-of-t-values} if the dimensionless quantity $t$
from Sect.~\ref{sec:Numerical-results} is identified with $t/b$ here.

The actual numerical value of $b$ in \eqref{eq:mod-FLRW}
is not really fixed by the matrix model
(a model with only dimensionless numbers), but may perhaps
be estimated~\cite{Klinkhamer2020}
as $b \sim \sqrt{\alpha^\prime}$
in terms of the Regge slope $\alpha^{\prime}$
from the inverse string tension of the
corresponding type--IIB superstring theory~\cite{IKKT-1997}.
Here, we have switched to natural units with $\hbar=1$  and $c=1$.

With the effective 
metric \eqref{eq:mod-FLRW}, the Ricci curvature scalar
$R(x) \equiv g^{\nu\sigma}(x)\,g^{\mu\rho}(x)\,R_{\mu\nu\rho\sigma}(x)$
and the Kretschmann curvature scalar
$K(x) \equiv R^{\mu\nu\rho\sigma}(x)\,R_{\mu\nu\rho\sigma}(x)$
are given by~\cite{Klinkhamer2019}
\begin{subequations}\label{eq:R-from-a-AND-K-from-a}
\begin{eqnarray}
\label{eq:R-from-a}
R\big[a(t)\big] &=&
\frac{6}{t^{3}}\,
\left(
t\,\big(b^{2} + t^{2}\big) \,\left(\frac{{a'(t)}}{{a(t)}}\right)^{2} +
\left[
t\,\big(b^{2} + t^{2}\big) \,\frac{a''(t)}{a(t)}
-b^{2}\,\frac{a'(t)}{a(t)}
\right]
\right) \,,
\end{eqnarray}
\begin{eqnarray}
\label{eq:K-from-a}
K\big[a(t)\big] &=&
\frac{12}{t^6}\,
\left( t^{2}\,{\big(b^{2} + t^{2}\big) }^{2}\,
\left(\frac{{a'(t)}}{{a(t)}}\right)^4
+
\left[
t\,\big(b^{2} + t^{2}\big) \,\frac{a''(t)}{a(t)}
-b^{2}\,\frac{a'(t)}{a(t)}
\right]^{2} \,\right) \,.
\end{eqnarray}
\end{subequations}
Both curvature scalars
are perfectly regular at $t=0$ for the series
$a(t)=1+a_2\,t^{2}/b^{2}+a_4\,t^4/b^4+ \ldots \;$;
see Ref.~\cite{Klinkhamer2020} for
further discussion.

Scalar metric perturbations of the background
\eqref{eq:mod-FLRW} have been studied
in Ref.~\cite{KlinkhamerWang2019}.
If nonrelativistic matter appears later (as argued in
Sect.~\ref{sec:Discussion}), the matter density perturbations
are found to grow with $|t|$ and it is possible to define a
``thermodynamic'' time $\mathcal{T}(t) = |t|$.
But, here, we only consider the emerging spacetime
without self-gravitating perturbations.

There is, however, an important caveat.
It is, namely, not at all clear that the genuine
Lorentzian large-$N$ matrix
model~\cite{IKKT-1997,Aoki-etal-review-1999,%
KimNishimuraTsuchiya2012,NishimuraTsuchiya2019},
and not just the set of corresponding classical
equations~\cite{Hatakeyama-etal2020},
also gives a nonsingular bouncing cosmology as suggested by
Fig.~3 of Ref.~\cite{Hatakeyama-etal2020} and the top-row panels
of Figs.~\ref{fig:fig1} and \ref{fig:fig3} here.
If a cosmic bounce does not result from the matrix model,
then the effective spacetime metric \eqref{eq:mod-FLRW-ds2}
only holds for $t > 0$,
perhaps with $b=t_\text{emergence} \sim \sqrt{\alpha^\prime}$.
The metric \eqref{eq:mod-FLRW-ds2} is then
equivalent~\cite{Klinkhamer2019,Klinkhamer2020}
to the spatially flat Robertson--Walker metric with cosmic-time 
coordinate $\tau \equiv \sqrt{b^{2}+t^{2}} > b=t_\text{emergence}\,$.

\section{Matrix-model configurations}
\label{sec:Matrix-model-configurations}

The authors of Ref.~\cite{Hatakeyama-etal2020}
have obtained numerical solutions of
the ``classical'' equations for the ten bosonic
$N \times N$ traceless Hermitian matrices $A^K$
of the Lorentzian IKKT matrix model~\cite{IKKT-1997}
(see Sect.~\ref{sec:Discussion} for 
an explanation of the quotation marks).
With a particular \textit{Ansatz} to implement
the $3 + 6$ split of spatial dimensions found
previously~\cite{KimNishimuraTsuchiya2012,NishimuraTsuchiya2019},
the classical results of Ref.~\cite{Hatakeyama-etal2020}
indicate the emergence of a cosmic-time coordinate
$t$ and a flat \mbox{3-space} expanding with $|t|$.
The results of the present article
support this interpretation if we compare the observables
from Ref.~\cite{Hatakeyama-etal2020}
with those obtained from having random
points in an expanding 3-dimensional cube or ball
embedded in $\mathbb{R}^{3}$ (Sects.~\ref{sec:Procedure} 
and \ref{sec:Numerical-results}).
In addition, we have proposed an effective metric to describe
this emerging spacetime (Sect.~\ref{sec:Effective-metric}).

At this point, it may be of interest to clarify the
meaning of the emerging ``time'' in the reduced classical version of the
large-$N$ matrix model~\cite{Hatakeyama-etal2020}.
From the \textit{Ansatz} (2.7)
of Ref.~\cite{Hatakeyama-etal2020} with $N=N_{X}\,N_{Y}$,
there are four $N_{X} \times N_{X}$ Hermitian matrices $X^\mu$,
for $\mu \in \{0,\, 1,\, 2,\, 3\}$. The particular matrix $X^0$,
singled out by the coupling constants $\eta_{KL}$,
is diagonalized and its real eigenvalues $\alpha_{p}$,
for $p \in \{1,\,2,\, \ldots\, ,N_{X}\}$, are ordered as follows:
\begin{equation}\label{eq:alpha-ordering}
\alpha_k \leq \alpha_{k+1}\,, \;\;\text{for}\;\;
k \in \{1,\,2,\,\ldots\, ,N_{X}-1\}\,.
\end{equation}
This essentially gives a dimensionless time variable $t$, which runs
parallel to the index $p$ of the eigenvalues $\alpha_{p}$,
as will become clear shortly.

In fact, definition (3.3) in Ref.~\cite{Hatakeyama-etal2020}
can be written as follows:
\begin{equation}\label{eq:t-k-def}
t_k \equiv
\frac{1}{n}\,\sum_{r=1}^{n} \alpha_{k+r}\,,
\end{equation}
with $n$ a fixed number ($1 \leq n < N_{X}$)
used for averaging the eigenvalues $\alpha_{p}$
and $k$ an index running over $\{0,\, 1,\, \ldots \,,\, N_{X}-n\}$.
In Ref.~\cite{Hatakeyama-etal2020} and the present article,
we simply write ``$t$'' for this emerging time,
\begin{equation}\label{eq:t=t-k}
t \equiv t_k \,.
\end{equation}

The results of Fig.~1 in Ref.~\cite{Hatakeyama-etal2020}
show that the distance between neighboring
eigenvalues $\alpha_p$ is more or less uniform.
Write $\Delta\alpha$
for this average distance, so that $\Delta\alpha=O(1/N_{X})$.
Then, we have approximately
\begin{equation}\label{eq:alpha-p-approximation}
\alpha_p \sim \alpha_1 + (p-1)\, \Delta\alpha\,,  
\end{equation}
for $p \in \{1,\,2,\, \ldots\, ,N_{X}\}$.
Assuming that $n$ is even, for simplicity,
definition \eqref{eq:t-k-def} gives with
\eqref{eq:t=t-k} and
\eqref{eq:alpha-p-approximation}
\begin{equation}\label{eq:t-from-alpha-p-approximation}
t \sim \alpha_{k+n/2} -\alpha_{N_{X}/2}\,,
\end{equation}
where the zero of $t$ has been defined to occur halfway,
at $k=N_{X}/2-n/2$. With fixed values of $n$ and $N_{X}$,
the increase of this dimensionless variable $t$
just follows the index $k$ on the $\alpha$ eigenvalues
in \eqref{eq:t-from-alpha-p-approximation}.

The crucial insight, now, is that \emph{time}
is really a tool for \emph{ordering} physical events.
And the $X^0$ eigenvalues $\alpha_p$ are ordered, 
according to \eqref{eq:alpha-ordering}.
Hence, the notion of time in $t_k$ of \eqref{eq:t-k-def}
is carried by the index $k$,
which corresponds to the index of the ordered eigenvalues
from Fig.~1 in Ref.~\cite{Hatakeyama-etal2020}.

We thus have an understanding of how the cosmic time $t$
emerges from the classical solutions of the large-$N$ matrix model.
In addition, the particular classical solutions
of Ref.~\cite{Hatakeyama-etal2020}
have an emerging flat 3-space, which has been found to expand
with $|t|$. The obvious question then is:
\emph{what drives this expansion?}
The answer appears to be: \emph{algebra}.
How precisely the algebra produces the expansion is not yet
completely clear, but we can sketch the beginning of an
explanation.

The results of Fig.~2 in Ref.~\cite{Hatakeyama-etal2020}
show that the spatial
matrices $X^{i}$ of the classical solution are nearly diagonal,
in the basis that makes $X^0$ perfectly diagonal 
(an ingredient for the definition of ``$t$'' as explained above). 
The effective size $R^{2}(t)$ of the universe is defined by
Eq.~(3.4) in Ref.~\cite{Hatakeyama-etal2020}
and shown by Fig.~3 in that reference.
This $R^{2}(t)$ behavior is more or less consistent with the behavior
on the diagonal $i=j$ of Fig.~2 in Ref.~\cite{Hatakeyama-etal2020}.
Let us look at the origin of these classical solutions
(or, rather, matrix-model configurations, as will be explained
in Sect.~\ref{sec:Discussion}).

The reduced classical equations are given by Eqs.~(2.16)
and (2.17) of Ref.~\cite{Hatakeyama-etal2020}.
For simplicity, we can take
a trivial solution $Y^a=0$ for the six internal-space
$N_{Y} \times N_{Y}$ Hermitian matrices
labeled by $a \in \{4,\, 5,\,\ldots \,,\, 9\}$ and
having dimension $N_{Y}=N/N_{X}$.
(The fermionic matrices $\Psi_\alpha$,
for $\alpha \in \{1,\, 2,\,\ldots \,,\, 16\}$, are also set to zero.)
We then focus on the four spacetime $N_{X} \times N_{X}$ Hermitian
matrices $X^\mu$ labeled by $\mu \in \{0,\, 1,\, 2,\, 3\}$,
for which the classical ``equations of motion'' are~\cite{Hatakeyama-etal2020}
\begin{subequations}\label{eq:X-mu-equations-of-motion}
\begin{eqnarray}
\delta_{ij}\,\Big[X^{i},\, \big[X^{j},\, X^0\big] \Big] - \xi\,X^0  &=& 0\,,
\\[2mm]
\eta_{\mu\nu}\,\Big[X^\mu,\, \big[X^\nu,\, X^{i}\big] \Big] - \zeta\,X^{i}  &=& 0\,,
\;\;\;\text{for}\;\;\; i \in \{1,\, 2,\, 3\} \,,
\end{eqnarray}
\end{subequations}
with Lagrange multipliers $\xi\ne 0$ and $\zeta\ne 0$.
We are only interested in nontrivial solutions $X^\mu\ne 0$.

Fixing the matrix size to $N_{X}=64$, for definiteness,
we can start with a diagonal block-type solution at $i \sim j \sim 32$
and then work out towards $i \sim j \sim 0$ and $i \sim j \sim 64$,
finding that the amplitudes of the matrices $X^{i}$
increase (cf. Fig.~2 in Ref.~\cite{Hatakeyama-etal2020}).
As the distance along the diagonal $i=j$ can be
interpreted as the cosmic time $t$
[roughly $t \sim (i-32)$ for $i=1,\, \ldots\,,\, 64$],
we have that the $X^{i}$ amplitudes grow with $|t|$,
i.e., the 3-space expands
(cf. Fig.~3 in Ref.~\cite{Hatakeyama-etal2020} and
the top-row panels of Fig.~\ref{fig:fig3} in the present article).
Further discussion is relegated to Appendix~\ref{app:Exact-solutions},
which presents certain exact solutions of the reduced classical equations
from the matrix model and makes a conjecture as to the
algebraic origin of the expansion.

\section{Discussion}
\label{sec:Discussion}

Heuristically, we now have the following description
of the origin of the Universe, within the context of
the IIB matrix model~\cite{IKKT-1997,Aoki-etal-review-1999}.
The fundamental matrix model has no spacetime and no gravity,
so that these physical quantities must somehow emerge from the
matrices of the model.

Certain ``classical'' matrix solutions were presented
in  Ref.~\cite{Hatakeyama-etal2020}, where
the quotation marks are there to alert us to the fact
that the IIB matrix model has no obvious small dimensionless 
parameter and, therefore, no obvious saddle-point approximation.
Hence, the matrices obtained are, at best, an approximation
to the dominant configurations of the path integral.
Still, these classical solutions
(or, rather, matrix-model configurations, as they are only
approximations) show a remarkable behavior 
that suggests an
emerging spacetime having a flat 3-space expanding with time
due to the algebra of the reduced equations of
motion \eqref{eq:X-mu-equations-of-motion}.

As mentioned above, gravitation and matter
are expected to appear dynamically.
For example, with ponderable matter present, the gravitational force
will be mediated by the appropriate massless modes
(virtual gravitons) of the emerging superstrings.
An interesting phenomenon will be the crossover 
from matrix-model expansion driven by the algebra
to Friedmann-type expansion driven by gravity and matter.
The latter type of expansion is, strictly speaking,
also due to (complicated) algebra, but that expansion can be
described by Einstein's general relativity as an effective theory.
It is possible that the emergent spacetime metric has the
form of 
that of the regularized big bang
singularity~\cite{Klinkhamer2019,Klinkhamer2020}, which is a
degenerate metric that solves the Einstein gravitational field equation
and has an external length parameter (now to be determined
by the matrix model).

However, at this moment, the urgent task is to show unequivocally that
the matrix model~\cite{IKKT-1997}
gives rise to the type of matrix-model configurations found
in Ref.~\cite{Hatakeyama-etal2020} and discussed in the present article.

\section*{Note added} 

The preprint version~\cite{Klinkhamer2019-v6} of the present article
contained an Appendix~B entitled ``IIB matrix model: Conceptual question,''
which discussed the possible role of the large-$N$ master field
for the emergence of a classical spacetime.
This appendix has, however, been removed in the published version,
as the content of the appendix was somewhat disconnected
from the rest of the article. 
Three follow-up articles on
the role of the IIB-matrix-model master field
for the emergent classical spacetime have appeared
subsequently~\cite{Klinkhamer2020a,Klinkhamer2020b,Klinkhamer2020c}.

\section*{Acknowledgments} 

We thank J.~Nishimura for useful comments
and acknowledge support by the KIT-Publication Fund
of the Karlsruhe Institute of Technology.

\appendix

\section{Exact solutions of reduced equations}
\label{app:Exact-solutions}

\subsection{General solutions}
\label{app:General-solutions}

In this appendix, we present certain exact solutions
of the reduced classical equations from Ref.~\cite{Hatakeyama-etal2020},
in order to get a better understanding of the numerical
solutions obtained in that reference. 
We first discuss the general solutions.

Specifically, we consider the
reduced classical equations \eqref{eq:X-mu-equations-of-motion}
for the four $N_{X} \times N_{X}$
Hermitian matrices $X^\mu$
with the following values of the Lagrange multipliers:
\begin{equation}
\label{eq:xi-and-zeta-equal-1}
\xi=\zeta=1\,,
\end{equation}
so that the set of four equations reads explicitly
\begin{subequations}\label{eq:X-mu-equations-of-motion-app}
\begin{eqnarray}
\sum_{j=1}^{3}\,\Big[X^{j},\, \big[X^{j},\, X^0\big] \Big] - X^0  &=& 0\,,
\\[2mm]
\sum_{j\ne i}\,\Big[X^{j},\, \big[X^{j},\, X^{i}\big] \Big]
-\Big[X^0,\, \big[X^0,\, X^{i}\big] \Big]
- X^{i}  &=& 0\,,
\;\;\;\text{for}\;\;\; i \in \{1,\, 2,\, 3\} \,.
\end{eqnarray}
\end{subequations}
This set of algebraic equations is surprisingly subtle. We have
two preliminary remarks. First, none of the matrices $X^\mu$
can be proportional to the identity matrix
(excluding the null matrix), as the commutator
terms in \eqref{eq:X-mu-equations-of-motion-app} then vanish
but the single-matrix terms not.
Second, the equations are nonlinear and allow
only for a trivial sign rescaling, $X^\mu \to  s^\mu\, X^\mu$, 
without summing over $\mu$
and with $s^\mu= \left( \pm 1,\,  \pm 1,\, \pm 1,\, \pm 1 \right)$.

We start by considering $2\times 2$ real symmetric matrices
$\widetilde{X}^\mu$, which will later be inserted on 
the diagonals of the respective $X^\mu$
matrices. An exact solution of the reduced classical equations
 \eqref{eq:X-mu-equations-of-motion-app},
restricted to $2\times 2$ traceless matrices, is given by:
\begin{subequations}\label{eq:X-mu-2by2-solution-general}
\begin{align}
\widetilde{X}^0 &= \left(
          \begin{array}{cc}
            \;\;\overline{a}_0\;\; & \;\;0\;\; \\
            0 & \;\;-\overline{a}_0\;\;  \\
          \end{array}
        \right)\,,
\hspace*{-30mm}
&\widetilde{X}^{2} &= \left(
          \begin{array}{cc}
            \;\;\overline{a}_2\;\; & \;\;\overline{b}_2\;\; \\
            \;\;\overline{b}_2\;\; & \;\;-\overline{a}_2\;\;  \\
          \end{array}
        \right)\,,
\\[2mm]
\widetilde{X}^{1} &= \left(
          \begin{array}{cc}
            \;\;a_1\;\; & \;\;b_1\;\; \\
            \;\;b_1\;\; & \;\;-a_1\;\;  \\
          \end{array}
        \right)\,,
\hspace*{-30mm}
&\widetilde{X}^{3} &= \left(
          \begin{array}{cc}
            a_3 & b_3 \\
            \;\;b_3\;\; & \;\;-a_3\;\;  \\
          \end{array}
        \right)
\,,
\end{align}
with
\begin{eqnarray}
\hspace*{-8mm}
\label{eq:a0bar}
\overline{a}_0&=&
\frac{1}{2}\;
{\sqrt{\frac{
4\,{a_{1}}^{2}\,\left( 1 - 4\,{b_{3}}^{2} \right)
+ 4\,{a_{3}}^{2}\,\left( 1 - 4\,{b_{1}}^{2} \right)
+ 32\,a_{1}\,a_{3}\,b_{1}\,b_{3} -1 +4\,{b_{1}}^{2} + 4\,{b_{3}}^{2}}
{1 - 4\,{b_{1}}^{2} - 4\,{b_{3}}^{2}}}}\,,
\\[2mm]
\label{eq:a2bar}
\hspace*{-8mm}
\overline{a}_2&=&-2\;
\frac{ a_{1}\,b_{1} + a_{3}\,b_{3}}
     {{\sqrt{1 - 4\,{b_{1}}^{2} - 4\,{b_{3}}^{2}}}}
\,,
\\[2mm]
\label{eq:b2bar}
\hspace*{-8mm}
\overline{b}_2&=&
\frac{1}{2}\; {\sqrt{1 - 4\,{b_{1}}^{2} - 4\,{b_{3}}^{2}}}
\,.
\end{eqnarray}
\end{subequations}
Hence, we have a solution $\widetilde{X}^\mu$ with
four moduli $\{ a_1,\,  b_1,\, a_3,\, b_3 \}$
that lie in a particular domain $D$,
so that the other three coefficients  are real:
\begin{equation}
\label{eq:domain-D}
D =
\left\{
\left( a_1,\,  b_1,\, a_3,\, b_3 \right) \in \mathbb{R}^4
\;\Big|\;
 \overline{a}_0\in \mathbb{R} \,\wedge\,
 \overline{a}_2\in \mathbb{R} \,\wedge\,
 \overline{b}_2\in \mathbb{R}
\right\}\,.
\end{equation}
We now establish that $D$ is nonempty.

Consider a hypercube $C_4$ in moduli space with the center at
\begin{subequations}\label{eq:C4}
\begin{eqnarray}\label{eq:C4-center}
\{ a_1,\,  b_1,\, a_3,\, b_3 \}\,\Big|^{(C_4-\text{center})} =
\{ 1,\,  1/4,\,1,\, 1/4 \}
\end{eqnarray}
and corners at
\label{eq:C4-corners}
\begin{eqnarray}
\{ a_1,\,  b_1,\, a_3,\, b_3 \}\,\Big|^{(C_4-\text{corners})} =
\{ 1\pm 1/8,\,  1/4\pm 1/16,\,1\pm 1/8,\, 1/4\pm 1/16 \}\,.
\end{eqnarray}
\end{subequations}
It is then easy to verify that moduli inside this hypercube $C_4$
give real values for $\{\overline{a}_0,\, \overline{a}_2,\, \overline{b}_2\}$, and the conclusion is that
\begin{equation}
\label{eq:C4-results}
C_4 \ne \varnothing  \wedge
C_4 \subset D \subset  \mathbb{R}^4\,.
\end{equation}
Explicitly, for the moduli \eqref{eq:C4-center} corresponding
to the center of the hypercube $C_4$ in moduli space,
we have the following solution:
\begin{subequations}\label{eq:X-mu-2by2-solution-center-C4}
\begin{align}
\widetilde{X}^0 &= \left(
          \begin{array}{cc}
            \;\;\sqrt{15}/2\;\; & 0 \\[1mm]
            0 & \;\;-\sqrt{15}/2\;\;  \\
          \end{array}
        \right)\,,
\hspace*{-0mm}             
&\widetilde{X}^{2} &= \left(
          \begin{array}{cc}
            \;\;-\sqrt{2}\;\; & \;\;1/(2\sqrt{2})\;\; \\[1mm]
            1/(2\sqrt{2}) & \sqrt{2}  \\
          \end{array}
        \right)\,,
\\[2mm]
\widetilde{X}^{1} &= \left(
          \begin{array}{cc}
            \;\;1\;\; & \;\;1/4\;\; \\[1mm]
            1/4 & -1  \\
          \end{array}
        \right)\,,
\hspace*{-0mm}
&\widetilde{X}^{3} &= \left(
          \begin{array}{cc}
            \;\;1\;\; & \;\;1/4\;\; \\[1mm]
            1/4 & -1  \\
          \end{array}
        \right)
\,,
\end{align}
\end{subequations}
with nonvanishing commutators $[\widetilde{X}^\mu,\,\widetilde{X}^\nu]$
for $(\mu,\,\nu)=$  $(0,\,1)$, $(0,\,2)$, $(0,\,3)$,
$(1,\,2)$, and $(2,\,3)$.

As to the actual shape of the 
manifold $D$ from \eqref{eq:domain-D}, we have three simple remarks.
First, we see from the root in \eqref{eq:b2bar}
that the moduli $(b_1 ,\,b_3)$ lie on a disk of
radius $1/(2\sqrt{2})$.
Second, we see from the numerator of the root in \eqref{eq:a0bar}
that the point $(a_1 ,\,a_3)=(0 ,\,0)$ is excluded and that
$|a_1|$ and $|a_3|$ must lie
above a boundary value determined by $(b_1 ,\,b_3)$.
Third, we also see from the numerator of the root in \eqref{eq:a0bar}
that the moduli $a_1$ and $a_3$ can run off to $\pm\infty$ and we
conclude that $D$ is a noncompact manifold.

With the $\widetilde{X}^\mu$
solution \eqref{eq:X-mu-2by2-solution-general} in hand,
it is straightforward to construct $N_{X}\times N_{X}$ traceless
Hermitian matrix solutions $X^\mu$ of the algebraic
equations \eqref{eq:X-mu-equations-of-motion-app}.
Taking
\begin{subequations}\label{eq:NX-2k-Xmu-block-sol}
\begin{eqnarray}\label{eq:NX-2k}
N_{X} &=& 2\,k\,, \;\;\text{for}\;\; k \in \mathbb{N}_{+}\,,
\end{eqnarray}
we have the following block-diagonal solutions:
\begin{eqnarray}
\label{eq:Xmu-block-sol}
X^\mu &=&
\text{diag}
\left(\widetilde{X}^\mu_{(1)},\,  \widetilde{X}^\mu_{(2)},\,
\ldots \, ,\,\widetilde{X}^\mu_{(k)}\right)\,,
\end{eqnarray}
\end{subequations}
where the suffixes $(1)$,\, $(2)$,\, \ldots\,,\,$(k)$
remind us that the individual blocks may have different values of the
moduli, as long as these moduli remain
in the domain $D$ from \eqref{eq:domain-D}.

At this moment, it may be of interest to compare our
exact solutions \eqref{eq:NX-2k-Xmu-block-sol} with the earlier
exact solutions of
Refs.~\cite{KimNishimuraTsuchiya2012-PRD,KimNishimuraTsuchiya2012-JHEP}.
Apart from one exception (mentioned in App.~C
of Ref.~\cite{KimNishimuraTsuchiya2012-JHEP}),
these earlier solutions have the property
that \emph{all} spatial matrices commute between themselves.
That space--space commutativity property does not hold  for the solutions
presented here.  For completeness, we should also mention
that other space--space noncommutative solutions have been considered
before; see, e.g., Ref.~\cite{Steinacker2019} and references therein.

\subsection{Special solutions}
\label{app:Special-solutions}

We next investigate whether or not we can qualitatively
reproduce the numerical results from Ref.~\cite{Hatakeyama-etal2020}
with the exact solutions of Appendix~\ref{app:General-solutions}.

In order to simplify the discussion, we take
\begin{eqnarray}\label{eq:NX-4l-plus-2}
N_{X} &=& 4\,l+2\,, \;\;\text{for}\;\; l \in \mathbb{N}_{0}\,.
\end{eqnarray}
There are four ingredients in our construction of a special
type of solution:
\begin{enumerate}
  \item
starting from \eqref{eq:Xmu-block-sol},
assume that the moduli $\{ a_1,\,  b_1,\, a_3,\, b_3 \}$
of the different blocks
are even functions of a single dimensionless time variable $\tau$;
  \item
set $\tau=0$ for the midway block $\widetilde{X}^\mu_{(l+1)}$;
  \item
order the diagonal entries of $X^0$  by a suitable conjugation
transformation (see below);
  \item
perform the same conjugation transformation on the other three
matrices $X^{i}$.
\end{enumerate}
The conjugation transformation used in the last two
ingredients traces, in fact, back to
a global gauge transformation of the original
matrix-model variables~\cite{IKKT-1997,Hatakeyama-etal2020}
and is given by
\begin{equation}\label{eq:U-transformation}
X^\mu \to  \underline{X}^\mu = U \cdot X^\mu \cdot U^{\dagger}\,,
\end{equation}
where $U$ is a single unitary matrix (for us, a single orthogonal matrix).
With the transformed matrices $\underline{X}^\mu$,
we calculate the following observable~\cite{Hatakeyama-etal2020}:
\begin{equation}\label{eq:Delta-matrix-def}
\Delta_{pq} \equiv \sum_{i=1}^{3}
\big| \big(\underline{X}^{i}\big)_{pq} \big|^{2}\,,
\end{equation}
for indices $p,q \in \{1,\,2,\, \ldots\, ,N_{X}\}$.

As a start, we get results for $N_{X}=6$ (or $l=1$) and explain them in detail.
From the first two ingredients listed above,
we have the block-diagonal solutions
\begin{eqnarray}\label{eq:Xmu-block-sol-NX-is-6}
X^\mu &=&
\text{diag}
\left(\widetilde{X}^\mu_{(\tau=-1)},\,  \widetilde{X}^\mu_{(\tau=0)},\,
\widetilde{X}^\mu_{(\tau=1)}\right)\,,
\end{eqnarray}
where the $2\times 2$ block $\widetilde{X}^\mu$
is given by \eqref{eq:X-mu-2by2-solution-general}
and $\tau$ labels the moduli of the different blocks.
In fact, we take the following functions
for the moduli:
\begin{subequations}\label{eq:moduli-functions-NX-is-6-second-choice}
\begin{eqnarray}
a_1(\tau)&=& a_3(\tau) = 3/8+\tau^{2}/4\,,
\\[2mm]
b_1(\tau)&=& b_3(\tau) = 1/8\,,
\end{eqnarray}
\end{subequations}
with $\tau\in [-1,\,1]$.
From these moduli, the three other coefficients
entering the $2\times 2$ matrices \eqref{eq:X-mu-2by2-solution-general}
are:
\begin{subequations}\label{eq:a0bar-a2bar-b2bar-NX-is-6-second-choice}
\begin{eqnarray}
\overline{a}_0(\tau)
&=&
{\sqrt{2/7}}\;{\sqrt{
      8\,{\big(3/8 + \tau^{2}/4\big)}^{2} - 7/8 }}\,,
\\[2mm]
\overline{a}_2(\tau)
&=&
-{\sqrt{2/7}}\;\big(3/8 + \tau^{2}/4\big) \,,
\\[2mm]
\overline{b}_2(\tau)
&=&\frac{1}{4}\;\sqrt{7/2}\,.
\end{eqnarray}
\end{subequations}

\begin{figure}[t]
\begin{center} 
\includegraphics[width=1.00\textwidth]{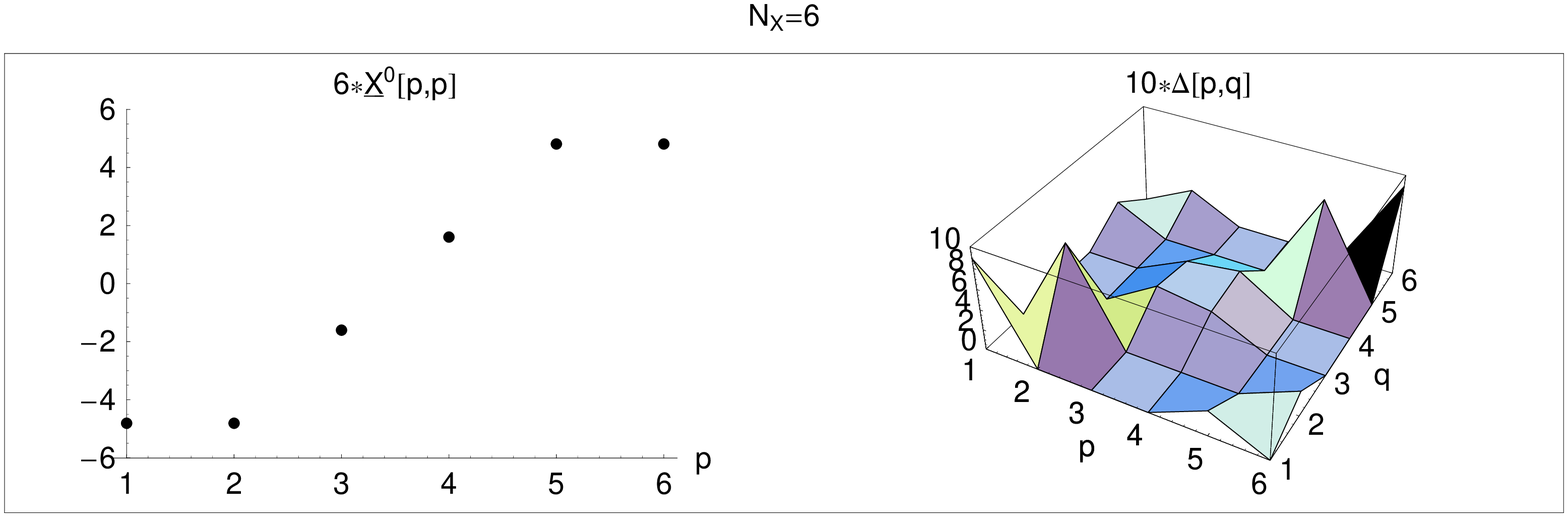}
\end{center}
\caption{Exact solution \eqref{eq:Xmu-block-sol-NX-is-6} for $N_{X}=6$ with
moduli \eqref{eq:moduli-functions-NX-is-6-second-choice}:
shown are, on the left,
the eigenvalues of the diagonal matrix $\underline{X}^0$
and, on the right,
the matrix $\Delta_{pq}$ defined by \eqref{eq:Delta-matrix-def}.
}
\label{fig:fig6}
\vspace*{0mm}
\vspace*{4mm}
\begin{center} 
\includegraphics[width=1.00\textwidth]{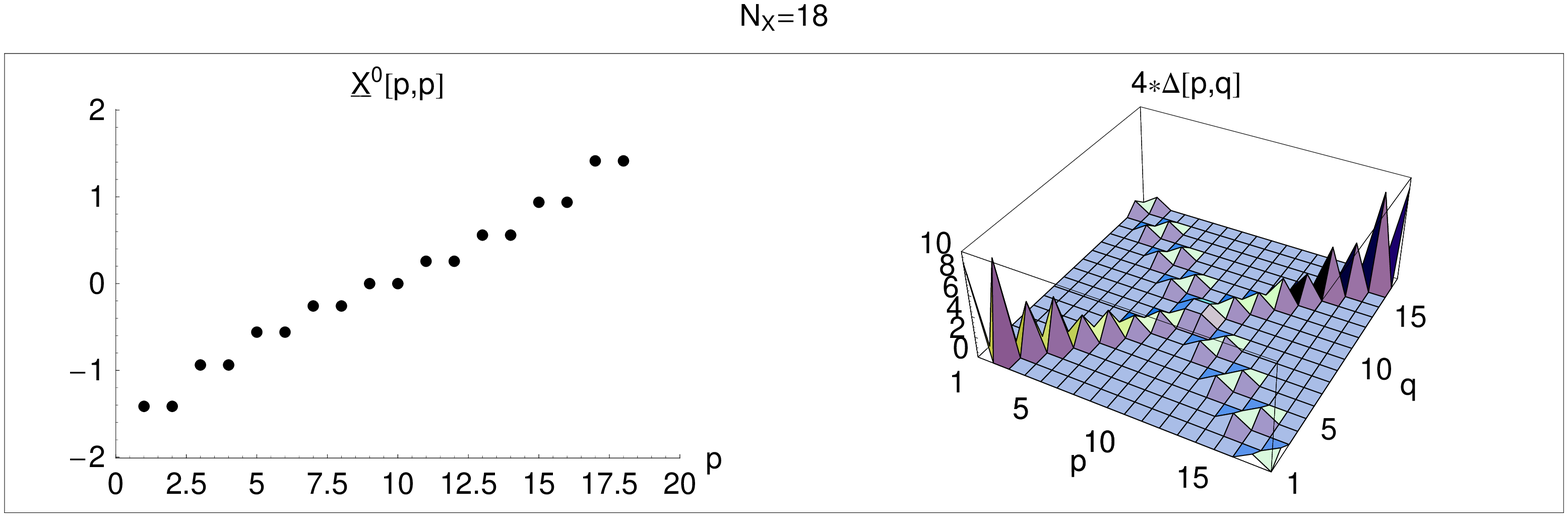}
\end{center}
\caption{Exact solution \eqref{eq:Xmu-block-sol-NX-general}
for $N_{X}=18$ with moduli \eqref{eq:moduli-functions-NX-general}:
shown are, on the left,
the eigenvalues of the diagonal matrix $\underline{X}^0$
and, on the right,
the matrix $\Delta_{pq}$ defined by \eqref{eq:Delta-matrix-def}.
}
\label{fig:fig7}
\end{figure}

Implementing the last two ingredients listed above,
the diagonal $X^0$ matrix from \eqref{eq:Xmu-block-sol-NX-is-6}
with moduli \eqref{eq:moduli-functions-NX-is-6-second-choice}
is reordered by use of the transformation \eqref{eq:U-transformation}
and the $X^{i}$ matrices are transformed simultaneously.
We then get the following values for the diagonal matrix $\underline{X}^0$
and the matrix $\Delta$ from \eqref{eq:Delta-matrix-def}:%
\begin{subequations}\label{eq:X0bar-Delta-results-NX-is-6-second-choice}
\begin{eqnarray}
\hspace*{-7mm}
\underline{X}^0 &=&
\text{diag}
\Big(-3/\sqrt{14},\ -3/\sqrt{14},\, -1/\sqrt{14},\, 1/\sqrt{14} ,\,
3/\sqrt{14},\,3/\sqrt{14}\Big)\,,
\\[2mm]
\hspace*{-7mm}
\Delta_{pq}  &=&
\left(
  \begin{array}{cccccc}
25/28 & 0  & 0  & 0  & 1/4 & 0  \\
0  &    25/28 & 0  & 0  & 0  & 1/4 \\
0  & 0  &  9/28 & 1/4 & 0  & 0 \\
0  & 0  & 1/4 & 9/28 & 0  & 0  \\
1/4 & 0  & 0  & 0  &  25/28 & 0 \\
0  & 1/4 & 0  & 0  & 0 & 25/28\\
\end{array}
\right)_{pq}\,,
\end{eqnarray}
\end{subequations}
which are also shown in Fig.~\ref{fig:fig6}.
The growth along the diagonal of $\Delta_{pq}$ is significant
(by a factor of approximately $3$, going from the middle of the
diagonal towards the edges) and there is a modest jump at $\tau=0$
in the $\underline{X}^0$ eigenvalues,
but the diagonal band structure of $\Delta_{pq}$ is
somewhat disturbed by small entries straddling the anti-diagonal.

It is now straightforward to generalize the $N_{X}=6$
construction to larger values of $N_{X}$.
With $l>1$ defined by \eqref{eq:NX-4l-plus-2},
the block-diagonal solutions are given by
\begin{eqnarray}
\label{eq:Xmu-block-sol-NX-general}
\hspace*{-16mm}
&&X^\mu =
\nonumber\\[1mm]
\hspace*{-16mm}&&
\text{diag}
\left(\widetilde{X}^\mu_{(\tau=-1)},
\, \ldots\, ,\,
\widetilde{X}^\mu_{(\tau=-2/l)},\,
\widetilde{X}^\mu_{(\tau=-1/l)},\,
\widetilde{X}^\mu_{(\tau=0)},\,
\widetilde{X}^\mu_{(\tau=1/l)},\,
\widetilde{X}^\mu_{(\tau=2/l)},\,
\ldots ,\,
\widetilde{X}^\mu_{(\tau=1)}
\right)\,.
\end{eqnarray}
We also make a slight change in the moduli functions
(the reason for this will become clear shortly),%
\begin{subequations}\label{eq:moduli-functions-NX-general}
\begin{eqnarray}
a_1(\tau)&=& a_3(\tau) = 1/4+\tau^{2}/2\,,
\\[2mm]
b_1(\tau)&=& b_3(\tau) = 1/4\,,
\end{eqnarray}
\end{subequations}
with $\tau\in [-1,\,1]$.
From these moduli, the three other coefficients
entering the $2\times 2$ matrices \eqref{eq:X-mu-2by2-solution-general}
are:
\begin{subequations}\label{eq:a0bar-a2bar-b2bar-NX-general}
\begin{eqnarray}
\overline{a}_0(\tau)
&=&
\sqrt{\tau^{2} + \tau^4}
\,,
\\[2mm]
\overline{a}_2(\tau)
&=&
- {\sqrt{2}}\;\big(1/4 +\tau^{2}/2\big)   \,,
\\[2mm]
\overline{b}_2(\tau)
&=&
\frac{1}{2\,{\sqrt{2}}}\,,
\end{eqnarray}
\end{subequations}
with $\overline{a}_0(0)=0$, as desired
[this is, in fact, the reason for having changed the moduli functions].
Reordering the diagonal $X^0$ matrix
from \eqref{eq:Xmu-block-sol-NX-general}
with moduli \eqref{eq:moduli-functions-NX-general}
and transforming the $X^{i}$ simultaneously,
we get the diagonal matrix $\underline{X}^0$ and the
matrix $\Delta$ from \eqref{eq:Delta-matrix-def}.
The results for $N_{X}=18$ are shown in Fig.~\ref{fig:fig7}.
Generalization to even
larger values of $N_{X}$ is trivial and the results will
look like those of  Fig.~\ref{fig:fig7}.

The exact results from
Fig.~\ref{fig:fig7} are qualitatively similar to the
numerical results of Figs.~1 and 2 in Ref.~\cite{Hatakeyama-etal2020}.
Admittedly, the right panel of our Fig.~\ref{fig:fig7}
has some noise along the anti-diagonal, but the main
feature of $\Delta_{pq}$ is the growth on the diagonal
(by a factor of approximately $9$, going from the middle of the diagonal
towards the edges).

It still needs to be explained how the particular moduli
functions \eqref{eq:moduli-functions-NX-general}
for the block-diagonal solution
\eqref{eq:Xmu-block-sol-NX-general} come about.
One conjecture is that there would be an ``entropy effect''
operative in the numerical procedure of Ref.~\cite{Hatakeyama-etal2020}:
the entropy for embedded block-diagonal solutions
with \mbox{$\tau$-dependent} moduli would be larger than
the entropy for embedded block-diagonal solutions
with constant ($\tau$-independent) moduli.

\end{document}